\newcommand{\nn}{\nonumber\\}

\newcommand{\be}{\begin{eqnarray}}
\newcommand{\ee}{\end{eqnarray}}

\newcommand{\beas}{\begin{eqnarray*}}
\newcommand{\eeas}{\end{eqnarray*}}
\def\gsim{\mathrel{\rlap{\lower4pt\hbox{\hskip1pt$\sim$}}
    \raise1pt\hbox{$>$}}}                
\documentclass[twocolumn,aps,showpacs]{revtex4}
\usepackage{color}
\usepackage{graphics}
\usepackage{epsfig}
\usepackage{amsmath}
\usepackage{amssymb}
\usepackage{amsthm}
\usepackage{amsbsy}
\begin{document}
\title{Spin alignment of vector mesons in heavy ion and proton - proton
  collisions}
\author{Alejandro Ayala$^{1,2}$\footnote{ayala@nucleares.unam.mx}, Eleazar
  Cuautle$^1$, G. Herrera Corral$^{3,4}$\footnote{On leave of absence from
  CINVESTAV}, 
  J. Magnin$^2$ and Luis Manuel Monta\~no$^4$}   
\affiliation{$^1$Instituto de Ciencias Nucleares Universidad
Nacional Aut\'onoma de M\'exico, Apartado Postal 70-543 M\'exico
Distrito Federal 04510, M\'exico.\\
$^2$Centro Brasileiro de Pesquisas F\1sicas CBPF,
Rua Dr. Xavier Sigaud 150 22290-180 Rio de Janeiro, Brazil.\\
$^3$CERN, CH 1211 Geneva 23, Switzerland.\\
$^4$Centro de Investigaci\'on y de Estudios Avanzados del IPN,
Apartado Postal 14-740 M\'exico Distrito Federal 07000, M\'exico.}

\begin{abstract} 

The spin alignment matrix element $\rho_{00}$ for the vector mesons $K^{*0}$
and $\phi(1020)$ has been measured in RHIC at central rapidities. These
measurements are consistent with the absence of polarization
with respect to the reaction plane in mid-central Au + Au collisions whereas,
when measured with respect to the production plane in the same reactions and
in p + p collisions, a non-vanishing and $p_\perp$-dependent $\rho_{00}$ is
found. We show that this behavior can be understood in a simple model of
vector meson production where the spin of their constituent quarks is oriented
during hadronization as the result of Thomas precession. 

\end{abstract}


\maketitle

\date{\today}

The study of spin polarization of produced hadrons in reactions at high
energies has opened a window to the understanding of the underlying dynamics
of quark recombination. In the context of heavy-ion collisions, polarization
studies can also help to understand the evolution of the system from its early
stages~\cite{Wang,Betz,Becattini}. 

Polarization analyses require to determine a given direction that serves 
as the spin quantization axis. From the experimental point of view, it is 
possible to determine two directions: the normal to the reaction and the 
normal to the production planes. The fist plane is defined as the one 
containing the impact parameter and the beam direction vectors whereas the 
second one is defined as containing the hadron's final momentum and the 
beam direction vectors.

Recently, the STAR collaboration has reported measurements of the spin 
polarization properties of the vector mesons $\phi$ and $K^*$~\cite{STAR}. 
These measurements refer to the $00$ component of the so called {\it spin 
alignment} density matrix $\rho$, which is the density matrix for a 
two-spin one-half system in a triplet state, expressed in terms of the 
coupled basis~\cite{Schilling}. Recall that a value $\rho^{00}=1/3$, means 
that the spin of the vector meson is not aligned with respect to the 
chosen quantization axis. Deviations from this value indicate a degree of 
polarization of the vector spin which ultimately might reflect a 
polarization of the constituent quarks.

The experimental findings reported can be summarized as follows: When the 
spin alignment is referred to the {\it reaction plane} in Au + Au 
collisions at $\sqrt{s_{NN}}=200$ GeV and measured at mid-rapidity, 
$\rho_{00}\simeq 1/3$, and it remains constant both as a function of 
$p_\perp$ in the range $0<p_\perp<5$ GeV, for mid-central collisions, and 
as a function of the average number of participants in the same $p_\perp$ 
range, for both $\phi$ and $K^*$. When the spin alignment is referred to 
the {\it production plane} and measured at mid-rapidity, both for p + p 
and mid-central Au + Au collisions at $\sqrt{s_{NN}}=200$ GeV, 
$\rho_{00}>1/3$ and it has a concave shape as a function of $p_\perp$ in the 
range $0<p_\perp<5$ GeV with minima at slightly different intermediate 
values of $p_\perp$ for $\phi$ and $K^*$.

An interesting observation that can be inferred from the above listed results
is that the dynamics of hadron formation seems to play a role to orient the
spin of quarks that form meson. 

A possible origin of a quark polarization driving the polarization of
vector mesons is discussed in Refs.~\cite{Wang} which study the transfer of
local relative angular momentum in peripheral nuclear collisions to the quark
spin polarization by means of rescattering during the reaction. However, this
mechanism also predicts a small global polarization growing almost linearly
with impact parameter which, according to the aforementioned results is not
observed in data.

Another interesting possibility that has only been studied in the context 
of hyperon polarization~\cite{DeGrand, Ayala} is the scenario where the 
spin of a quark is oriented during the recombination process. The 
semiclassical picture accounting for the polarization is the Thomas 
precession produced by the accelerating force that pulls a slow moving 
quark ($q^s$) to form a fast moving hadron~\cite{DeGrand}. This mechanism 
also predicts that if the quark is fast ($q^f$) and is decelerated to form 
the hadron, its polarization will be of opposite sign compared to the case 
when it is accelerated. The pulling force is required to not be parallel 
to the original quark velocity since the Thomas frequency is a vector 
formed by the cross product of the force ${\mbox{\bf F}}$ and this 
velocity $\boldsymbol\beta$, namely \be
   \boldsymbol\omega_T=\left(\frac{\gamma}{1+\gamma}\right){\mbox{\bf F}}
   \times\boldsymbol\beta,
\label{Thomasomega}
\ee
where $\gamma$ is the Lorentz gamma-factor. The polarization is given
by~\cite{DeGrand}
\be
   {\mathcal{P}^{s,f}}&=\mp&\frac{\omega_T^{s,f}}{\Delta E},
\label{polgeneral}
\ee
where the $-$ ($+$) sign refers to the $q^s$ ($q^f$). $\omega_T^{s,f}$ is the
magnitude of the Thomas precession frequency for $q^s$ and $q^f$, respectively
and $\Delta E$ is the change of energy in the process of hadron formation.
 
In this work we use the Thomas spin precession mechanism to describe the spin
alignment of vector mesons produced at central rapidity in Au + Au and p + p
collisions at $\sqrt{s_{NN}}=200$. We show that under very simple assumptions,
data for $\rho_{00}$ are well reproduced within this approach.

The physical picture we use is that of a fast quark that decelerates
and a slow one that accelerates to form a fast moving hadron. In the
process, Thomas precession makes the spin of the former to acquire a positive
polarization whereas the latter acquires a negative one. For central
rapidities, the large momentum component of the hadron will thus be its
transverse momentum $p_{\perp}^H$ whereas the small component will be its
longitudinal one, $p_\parallel^H$. We will assume that in the beam collision,
a hard interaction produces a fast quark moving with a large transverse
momentum $p_\perp^f$ and, to simplify matters, a vanishing longitudinal
momentum. This fast quark combines with the slow one, that we assume moves
originally mainly in the longitudinal direction with momentum $p_\parallel^s$
and, also for simplicity, take it with vanishing transverse momentum. This
sharp difference in the original direction of motion of the recombining quarks
is at the core of the produced polarization since, as we proceed to show, it
gives rise to a distinct $p_\perp$ dependence of $\rho_{00}$ which seems to be
also observed in data. 

In order to form the hadron, which should move with an intermediate value of
momentum, between that of the $q^f$ and of the $q^s$, the fast quark should
slow down whereas the slow quark should speed up. Notice that for this
mechanism to work, there is no need to assume that the process happens only
in either a proton-proton or a nucleus-nucleus collision. Notice also that the
momentum of the formed hadron provides a fixed direction to define
that $q^s$ ($q^f$) decelerates (accelerates), whereas, when referred to the
reaction plane, no such fixed direction exists, since the direction of the
impact parameter vector changes from one reaction to another and in such
situation either quark can accelerate or decelerate. 

The pulling force is equal to the change in momentum ${\bf \Delta p}$ of the
given quark, in the interval of time $\Delta t$ for the recombination process
to happen, that is 
\be
   {\mbox{\bf F}}=\frac{\bf \Delta p}{\Delta t}.
\label{force}
\ee
Thus $\omega_T$ for the given quark can be computed as the average over
this time interval~\cite{DeGrand}, namely
\be
   \omega_T^{s,f}&\propto&\frac{\Delta p^{s,f}}{\Delta t}\beta^{s,f}
   \left(\int_{\Delta t}dt\sin\theta^{s,f}/\Delta t\right)\nn
   &\simeq&\frac{\Delta p^{s,f}}{\Delta
     t}\beta^{s,f}\langle\sin\theta\rangle^{s,f}, 
\label{average}
\ee
where in the last line we have changed the time average
of the sine of the angle by the average sine of the angle
between the quark velocity vectors and their corresponding change in
momentum. From Eqs.~(\ref{polgeneral}) and~(\ref{average}), we see that
the calculation of the $q^f$ and $q^s$ polarizations reduces to computing the
magnitudes of their change in momentum $\Delta p^{s,f}$, their
$\langle\sin\theta\rangle^{s,f}$ and the change in energy $\Delta E$.

We first compute the change in momenta. For $q^s$, we have
\be
   \Delta p^s = \sqrt{(p_\parallel^{s/H} -
   p_\parallel^{s})^2 + (p_\perp^{s/H})^2},
\label{deltap}
\ee
where $p_{\parallel ,\ (\perp )}^{s/H}$ denote the parallel (transverse)
component of the $q^s$ in the hadron $H$. Let us assume that, in order to 
enhance the recombination probability, the rapidity of
$q^s$ has to be within the hadron's one. Under this assumption we can
write
\be
   p_\parallel^{s} &=& m_\perp^{s}\sinh y^H\nonumber\\
               &=& \frac{m_\perp^{s}}{m_\perp^H}
                   m_\perp^H\sinh y^H\nonumber\\
               &=& \frac{m^{s}}{m_\perp^H}p_\parallel^H,
\label{ppara}
\ee
where we have set $m_\perp^{s}=m^{s}$ since
$p_\perp^{s}=0$. Therefore, we can write
\be
   \Delta p^{s} &=& \sqrt{[(x_\parallel -
   \frac{m^s}{m_\perp^H})p_\parallel^H]^2 + [x_\perp
       p_\perp^H]^2}\nonumber\\
   &\simeq&x_\perp P_\perp^H,
\label{deltaapprox}
\ee
where we have introduced the definitions for the momentum fractions that the
$q^s$ has inside the hadron,
\be
   x_\parallel &=& p_\parallel^{s/H}/p_\parallel^H\nn
   x_\perp &=& p_\perp^{s/H}/p_\perp^H,
\label{xperpxpara}
\ee
and have neglected the longitudinal hadron's momentum with
respect to its transverse one. 

Similarly, for the $q^f$ change of momentum when decelerating, notice that we
have
\be
   {\bf \Delta p}^f = {\bf p}_\parallel^{f/H} -  {\bf p}_\perp^{f/H} -
   {\bf p}_\perp^{f}.
\label{Deltavec}
\ee
Since $\boldsymbol\beta^f$ initially points along the perpendicular direction 
and, although this velocity vector changes so that the final hadron's momentum
eventually picks up a longitudinal component, the dominant component of the
$q^f$ in the hadron is the transverse one. Therefore, for the cross product of 
${\bf \Delta p}^f$ with $\boldsymbol\beta^f$ we get 
\be
   {\bf\Delta p}^f\times\boldsymbol\beta^f
   &=&p_\parallel^{f/H}\beta^f\langle\sin\theta\rangle^f\nn
   &=&(1-x_\parallel)p_\parallel^H\langle\sin\theta\rangle^f ,
\label{deltaptimesbeta}
\ee
where we have enforced momentum conservation $x_\perp + x_\parallel = 1$ and
have approximated $\beta^f\simeq 1$. 

Now, we proceed to compute the average sine of the angles. 
To compute $\langle\sin\theta\rangle^s$, notice that given that initially
$\boldsymbol\beta^s$ points along the parallel direction and ${\bf\Delta p}^s$
is basically directed along the perpendicular direction, the initial angle
between these vectors is $\pi/2$ and the average one should be 
close to $\pi/4$. Thus $\langle\sin\theta\rangle^s\simeq
1/\sqrt{2}$. Rather than approximating $\beta^s$ (and therefore $\gamma^s$),
 we introduce a factor $a$ for this polarization and let this vary such that 
$0<a<1$. Thus we write
\be
\omega^s &=& a \frac{\Delta p^s}{\Delta t} \nonumber \\
a &=& \left(\frac{\gamma^s}{1+\gamma^s}\right)\beta^s\left<\sin\theta\right>^s.
\ee

To compute
$\langle\sin\theta\rangle^f$, notice that since ${\bf\Delta p}^f$ and
$\boldsymbol\beta^f$ are almost perpendicular,
$\langle\sin\theta\rangle^f\simeq 1$. 

The change in energy is common to both the accelerating $q^s$ and the
decelerating $q^f$
\be
   \Delta E &=& \{ [(p_\perp^f)^2 + (p_\parallel^f)^2 + (m^f)^2]^{1/2}\nn
   &+&
   [(p_\perp^s)^2 + (p_\parallel^s)^2 + (m^s)^2]^{1/2}\nn
   &-& [(p_\perp^H)^2 + (p_\parallel^H)^2 + (m^H)^2]^{1/2} \}\nonumber\\
   &\simeq& \{ [(p_\perp^f)^2 + (m^f)^2]^{1/2} +
   [(p_\parallel^s)^2 + (m^s)^2]^{1/2}\nn
   &-& [(p_\perp^H)^2 + (m^H)^2]^{1/2} \}\nonumber\\
   &=& \left\{ p_\perp^f \left[
   1+\frac{(m^f)^2}{2(p_\perp^f)^2}\right] +
   [(p_\parallel^s)^2 + (m^s)^2]^{1/2}\right.\nn 
   &-&\left.
   p_\perp^H \left[1+\frac{(m^H)^2}{2(p_\perp^H)^2}\right] 
   \right\},
\label{deltaEini}
\ee
where we have set $p_\perp^s=p_\parallel^f=0$, neglected
$(p_\parallel^H)^2$ compared with $(p_\perp^H)^2$ and expanded the square
roots assuming that the transverse momentum components are large. Introducing
the relation between the hadron and $q^f$ transverse momenta
$p_\perp^f=p_\perp^H/z$, with $0<z<1$, we get 
\begin{figure} 
{\centering
{\epsfig{file=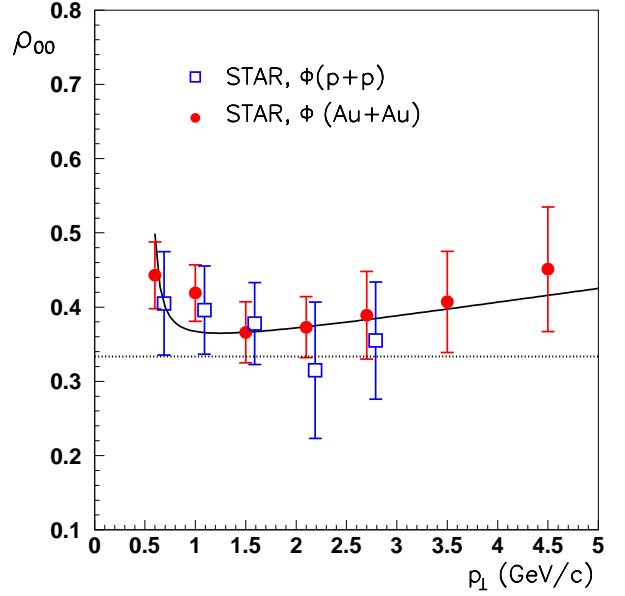, width=1\columnwidth}}
\par}
\caption{(Color online) $\rho_{00}$ as a function of $p_\perp$ for 
  $\phi$(1020) using the model parameters described in the text, 
  compared to data from STAR~\cite{STAR} for p +p and 
  Au + Au collisions at centrality 20-60
  \%, measured with respect to the production plane. 
  For clarity, the $p_\perp$ for p + p data has been displaced by 
  0.09 GeV with respect 
  to the reported central value. The statistical and systematic 
  errors have been added in quadrature. For comparison, we 
  also draw the constant value 1/3 that represents the absence of
  polarization.}  
\label{fig1}
\end{figure}
\be
   \Delta E &=& \left\{\frac{p_\perp^H}{z}\left( 1+ \frac{z^2(m^f)^2}
   {2(p_\perp^H)^2}  \right)\right.\nn 
   &+& \left[ 
   \left( \frac{m^s}{m_\perp^H}\right)^2(p_\parallel^H)^2
   + (m^s)^2  \right]^{1/2}\nn
   &-&\left. 
   p_\perp^H\left( 1+ \frac{(m^H)^2}{2(p_\perp^H)^2}  \right)\right\}
\label{deltaEcont}
\ee
where we have made use of the assumption that the rapidity of the $q^s$
coincides with that of the hadron. Notice that the above expression
can be simplified. In particular, since $p_\parallel^H=m_\perp^H\sinh y^H$, we
can write
\be
   \left[\frac{(m^s)^2}{(m_\perp^H)^2}(p_\parallel^H)^2 + 
   (m^s)^2\right]^{1/2}  = m^s\cosh y^H.
\label{aproxrap}
\ee
Therefore, $\Delta E$ can be expressed as
\be
  \Delta E &=& 
   \left\{ \left(\frac{1-z}{z}\right)p_\perp^H +
   \left[\frac{z(m^f)^2- (m^H)^2}{2p_\perp^H}\right]\right.\nn 
   &+&\left.
   m^s\cosh y^H
   \right\}.
\label{deltaEsimp}
\ee
We enphasize that the approximation to compute $\Delta E$ is such that
Eq.~(\ref{deltaEsimp}) is valid for $p_\perp\gsim m^H$ and that for $z\lesssim
1$ the validity of the approximation can be extended to lower values of
$p_\perp$. 

It is now easy to compute the polarization for the slow and fast quarks by
means of Eq.~(\ref{polgeneral}) and from them, the $\rho_{00}$ density matrix
element given by 
\be
   \rho_{00}=\frac{1-{\mathcal{P}}^{s}{\mathcal{P}}^{f}}
   {3+{\mathcal{P}}^{s}{\mathcal{P}}^{f}}.
\label{rho00}
\ee

Let us first proceed to apply the model to compute $\rho_{00}$ for
$\phi$. This case is the simplest one to treat within our approach since the
quark content of this particle is ${\mbox{s}}\bar{\mbox{s}}$ and thus either
one of these 
quarks can be thought of as being the fast ($q^f$) or the slow ($q^s$)
one. Rather than making an exhaustive search in the parameter space, we choose
reasonable values for them. We first fix the $\phi$ and strange quark masses
to be $m^\phi=1.02$ GeV, $m^{{\mbox{s}},\bar{\mbox{s}}}=0.5$ GeV. The rapidity
value we use is $y^H=1$ and the formation time $\Delta t=1$ fm. For the
fractions of longitudinal and transverse momenta that the slow quark has
inside the $\phi$ we take $x_\perp=x_\parallel=0.5$. The fraction of the
transverse momentum carried by the $\phi$ from the fast quark is taken as
$z=0.9$.

Figure~\ref{fig1} shows $\rho_{00}^\phi$ as a function of $p_\perp^\phi$ 
compared to data from STAR~\cite{STAR} for p + p and Au + Au collisions at 
centrality 20-60 \%. A good description is obtained for $a=0.25$.

We now proceed to apply this model to the case of $K^*$, whose quark content
is ${\mbox{d}}\bar{\mbox{s}}$. However, in this case one needs to be careful
since the symmetry between the masses, present in the description of $\phi$,
is absent. Consequently the spin alignment has to be treated in average. To
this end, let us first take a simple scenario and consider the arithmetic
average in the following manner
\be
   \rho_{00}^{K^*}=\frac{1}{2}\left(\rho_{00}^{f={\mbox{s}},\ s={\mbox{d}}}+
   \rho_{00}^{f={\mbox{d}},\ s={\mbox{s}}}\right).
\label{algebraicaverage}
\ee
\begin{figure} 
{\centering
{\epsfig{file=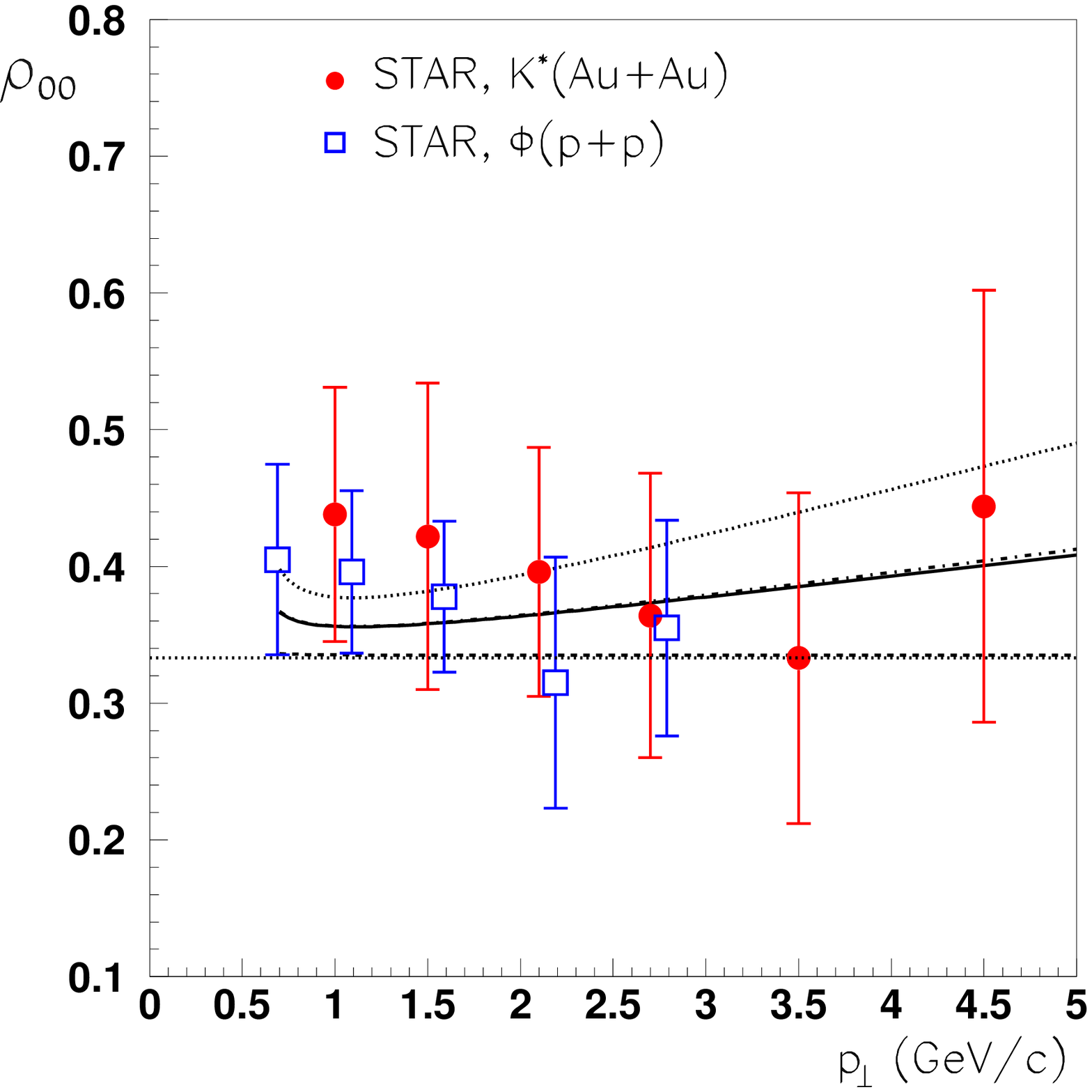, width=1\columnwidth}}
\par}
\caption{(Color online) $\rho_{00}$ as a function of $p_\perp$ for $K^*$ using
  the 
  model parameters described in the text, compared to
  data from STAR~\cite{STAR} for p +p and Au + Au collisions at centrality 
  20-60 \%, measured with respect to the production plane. For clarity, 
  the $p_\perp$ for p + p data has been displaced by 0.09 GeV with respect 
  to the reported central value. The statistical and 
  systematic errors have been added in quadrature. The upper 
  dashed curve represents the case
  for $\rho_{00}^{f={\mbox{s}},\ s={\mbox{d}}}$, whereas the lower dashed
  curve represents the case for $\rho_{00}^{f={\mbox{d}},\ s={\mbox{s}}}$. The
  intermediate solid curve represents $\rho_{00}^{K^*}$ as the algebraic
  average of the above. The intermediate dotted curve represents the case
  where $\rho_{00}$ is computed using the average product of
  polarizations. For comparison, we also draw the constant 
  value 1/3 that represents the absence of
  polarization.}
\label{fig2}
\end{figure}

Figure~\ref{fig2} shows $\rho_{00}^{K^*}$ as a function of $p_\perp^{K^*}$ 
compared to data from STAR~\cite{STAR} for p + p and Au + Au collisions at 
centrality 20-60\%, measured in the production plane. The curves are 
computed such that we employ the same set of parameters as in the 
computation of $\rho_{00}^\phi$ (with $m^{\mbox{d}}=0.3$ GeV) except for 
the value of $z$. The reason is that, whereas in the case that 
$q^f={\mbox{s}}$ one can think that in order for this fast quark to pick 
up a ${\mbox{d}}$, the momenta of ${\mbox{s}}$ and $K^*$ are similar, when 
$q^f={\mbox{d}}$, its momentum must be larger, given the mass difference 
between $K^*$ and ${\mbox{d}}$. Thus if $q^f={\mbox{s}}$ we choose $z=0.9$ 
whereas when $q^f={\mbox{d}}$ we use $z=0.3$. The upper dashed curve 
represents the case for $\rho_{00}^{f={\mbox{s}},\ s={\mbox{d}}}$, whereas 
the lower dashed curve is for $\rho_{00}^{f={\mbox{d}},\ s={\mbox{s}}}$. 
The intermediate solid curve represents $\rho_{00}^{K^*}$ as the algebraic 
average of the above, as in Eq.~(\ref{algebraicaverage}).

An alternative approach is to consider that the $\rho_{00}^{K^*}$ can be
computed by the substitution
\be 
   {\mathcal{P}}^{s}{\mathcal{P}}^{f}\rightarrow
   \frac{{\mathcal{P}}^{s={\mbox{s}}}{\mathcal{P}}^{f={\mbox{d}}}+
         {\mathcal{P}}^{s={\mbox{d}}}{\mathcal{P}}^{f={\mbox{s}}}}
   {2},
\label{sustitutingpol}
\ee
that is, by the {\it average product of polarizations}, in
Eq.~(\ref{rho00}). Figure~\ref{fig2} shows also this possibility 
represented by the intermediate dotted curve, using the same set of
parameters for the cases were $q^f={\mbox{s}}$, $q^s={\mbox{d}}$ and
$q^f={\mbox{d}}$, $q^s={\mbox{s}}$, as discussed above. As can be seen from
the figure, no significant difference is found with the case where the average
is taken with the functions $\rho_{00}$ and both approaches give a good
description of data.

In conclusion, we have shown that data on the vector mesons $\phi$ and $K^*$
spin alignment with respect to the production plane in Au + Au and p + p
collisions are well described by assuming that these hadrons are produced by
the recombination of a slow and a fast quark that in the process
become polarized in opposite directions due to Thomas precession. In this
plane, the momentum of the hadron provides a fixed direction to define whether
a valence quark accelerates or decelerates. The mechanisms also clarifies the
fact that when the spin alignment is referred to the
reaction plane, $\rho_{00}$ vanishes, given that no such fixed direction
exists, since impact parameter vector changes from one reaction to another. 

In the near future ALICE at the LHC will have the capability to measure and
reconstruct $\phi$ and $K^*$ mesons with larger statistics~\cite{alice}. In
addition, its particle identification will allow these meson's
$p_\perp$ to be measured beyond 5 GeV, well into the region where energy
losses become important and also where fragmentation (as opposed to
the recombination picture we are using here), becomes the
dominant particle production mechanism. It will thus be interesting to study
how the polarization of vector mesons changes when including these
effects. This is work for the future. 

\section*{Acknowledgements}

A.A. thanks the kind hospitality of faculty and staff members at CBPF during a
sabatical visit. Support for this work has been received in part by FAPERJ
under Proj. No. E-26/110.166/2009, CNPq, the Brazilian Council for Science and
Technology, PAPIIT-UNAM under grant Nos. IN116008 and IN116508 and by 
CONACyT-M\'exico.

\end{document}